\documentclass[useAMS,usenatbib]{mn2e}
\usepackage{mncite}
\usepackage{graphicx}
\usepackage{epstopdf}
\usepackage{subfigure}
\usepackage[a4paper]{hyperref}
\usepackage{amssymb}
\usepackage{aas_macros}
\usepackage[usenames]{color}

\title[The Doppler shadow of HD 189733b]{Line-profile tomography of
 exoplanet transits I: The Doppler shadow of HD 189733b
\thanks{Based on data collected with the HARPS spectrograph at ESO La Silla Observatory under the allocated programme 079.C-0828(A). The data are publicly available at the CDS ({\sc cdsarc.u-strasbg.fr}).}}
\author[A. Collier Cameron et al]
{
A. Collier Cameron$^{1}$\thanks{E-mail:acc4@st-and.ac.uk},
V. A. Bruce$^{1}$,
G. R. M. Miller$^{1}$,
A. H. M. J. Triaud$^{2}$ and
\newauthor
D. Queloz$^{2}$
\\
$^{1}$SUPA, School of Physics and Astronomy, University of St Andrews, North Haugh, St Andrews, Fife KY16 9SS, UK.\\
$^{2}$Observatoire de l'Universit\'e de Gen\`eve, Chemin des Maillettes 51, CH-1290 Sauverny, Switzerland.
}
\begin{document}

\date{Accepted 0000 December 00. Received 0000 December 00; in original form 0000 October 00}

\pagerange{\pageref{firstpage}--\pageref{lastpage}} \pubyear{2008}

\maketitle

\label{firstpage}

\begin{abstract}
The misalignment between the orbital plane of a transiting exoplanet and the spin axis of its host star provides important insights into the system's dynamical history. The amplitude and asymmetry of the radial-velocity distortion during a planetary transit (the Rossiter-McLaughlin effect) depend on the projected stellar rotation rate $v\sin I$ and misalignment angle $\lambda$, where the stellar rotation axis is inclined at angle $I$  to the line of sight. The parameters derived from modelling the R-M effect have, however, been found to be prone to systematic errors arising from the time-variable asymmetry of the stellar spectral lines during transit.
Here we present a direct method for isolating the component of the starlight blocked by a planet as it transits the host star, and apply it to spectra of the bright transiting planet HD 189733b. 
We model the global shape of the stellar cross-correlation function as the convolution of a limb-darkened rotation profile and a gaussian representing the Doppler core of the average photospheric line profile. The light blocked by the planet during the transit is a gaussian of the same intrinsic width, whose trajectory across the line profile yields a precise measure of the misalignment angle and an independent measure of $v\sin I$. 
We show that even when $v\sin I$ is less than the width of the intrinsic line profile, the travelling Doppler ``shadow'' cast by the planet creates an identifiable distortion in the line profiles which is amenable to direct modelling. Direct measurement of the trajectory of the missing starlight yields self-consistent measures of the projected stellar rotation rate, the intrinsic  width of the mean local photospheric line profile, the projected spin-orbit misalignment angle, and the system's centre-of-mass velocity. Combined with the photometric rotation period, the results give a geometrical measure of the stellar radius which agrees closely with values obtained from high-precision transit photometry if a small amount of differential rotation is present in the stellar photosphere.
\end{abstract}


\section{Introduction}

The recent discoveries of strong projected spin-orbit misalignments in the transiting exoplanets XO-3b (\citealt{hebrard2008_xo3_rm}, \citealt{winn2009xo3_rm}), HD 80606b (\citealt{winn2009hd80606rm}, \citealt{pont2009hd80606rm}) and WASP-14b \citep{johnson2009wasp-14rm}, and retrograde orbital motion in WASP-17b \citep{anderson2009wasp-17} and HAT-P-7b (\citealt{winn2009hat-p-7rm}; \citealt{narita2009hat-p-7rm}) indicate violent dynamical histories for a significant fraction of the population of hot-Jupiter planets. 

These misalignments are measured by obtaining densely-sampled radial-velocity observations during transits. The starlight blocked by the planet during a transit possesses the radial velocity of the obscured part of the stellar surface. The radial-velocity centroid of the light emanating from the visible parts of the stellar disk exhibits a reflex motion disturbance --the Rossiter-McLaughlin (RM) effect -- whose form depends on the projected stellar equatorial velocity  $v\sin I$, the impact parameter $b$ of the planet's path across the stellar disc, the projected misalignment angle $\lambda$ between the orbital axis and the stellar spin axis, and the relative sizes of the star and planet. 

Detailed semi-analytic formulations for determining the anomalous departure of the line centroid from the stellar reflex orbit have been developed by \citet{ohta2005rm}, \citet{gimenez2006rm} and \citet{hirano2009rm}. 
The process of measuring and interpreting this reflex velocity shift is not, however, straightforward. If the stellar rotation profile is even partially resolved by the spectrograph, the missing light within the planet's silhouette introduces an asymmetry into the line spread function. For instruments such as SOPHIE and HARPS, radial velocities are derived using a gaussian fit to a cross-correlation function (CCF) computed from the stellar spectrum and a weighted line mask (\citealt{baranne96elodie}; \citealt{pepe2002harps}). Such methods work extremely well for the symmetric and time-invariant profile shapes encountered outside transit. When applied to profiles with a time-varying degree of intrinsic asymmetry, however, they yield velocity measures that depart systematically from the velocity centroid of the visible starlight. 

\citet{winn2005rm} overcame this difficulty for their iodine-cell observations of HD 209458b using a direct modelling approach which they have employed in subsequent papers. They built models of both the unobscured stellar profile and the light blocked by the planet into their analysis of the line-spread function,  deriving semi-empirical corrections to the model velocities to enable meaningful comparison with the data. Nonetheless, their study of the RM effect in HD 189733 \citep{winn2006hd189733} shows a clear pattern of correlated radial-velocity residuals during the transit. In a more recent HARPS study of the RM effect during a transit of HD 189733b, \citet{triaud2009rm} noted an almost identical systematic pattern of correlated residuals between the radial velocities of the gaussian fits to the CCFs, and the line centroid velocities of the best-fitting model computed using the  formulation of \citet{gimenez2006rm}. 

In developing an analytic model of this velocity anomaly for cross-correlation spectra, \citet{hirano2009rm} have shown that
such errors are exacerbated as $v\sin I$ increases and the asymmetry becomes more extreme. If uncorrected, the anomaly can lead to significant over-estimation of the value of $v\sin I$, particularly for rapidly-rotating planet-host stars. The motivation for the present study is to develop a comprehensive model of  the changes in CCF  morphology that occur before, during and after a transit, in terms of  the projected stellar equatorial rotation speed, the spin-orbit misalignment angle, the intrinsic width of the combined stellar CCF and instrumental broadening function, and the stellar limb-darkening coefficient. In principle this approach is similar  to that of \citet{winn2005rm}, but we use it to decompose the observed CCF into its various components and track them directly. 
A similar methodology has already been used to model the RM effect in the binary stars V1143 Cyg \citep{albrecht2007} and DI Her \citet{albrecht2009}.  To illustrate the effectiveness of this approach as an alternative to gaussian-fitting in the study of transiting exoplanets, we re-analyse the same HARPS observations of HD 189733b described by \citet{triaud2009rm}. We show that the spectral signature of the light blocked by the planet is clearly discernible in the residuals when a model of the unobscured starlight is subtracted from the data, and track it directly to derive new measurements of the projected stellar rotation velocity and spin-orbit misalignment angle.

\section{Observations and data analysis}
\label{sec:obsred}

Densely-sampled spectral time-series of HD 189733 were obtained with the HARPS instrument on the ESO 3.6-m telescope at La Silla on the nights of 2006 September 8 and 2007 August 29. The observations are described in detail by \citet{triaud2009rm}. The 2006 September data were taken with a cadence of 10.5 minutes, while the 2007 data were sampled every 5.5 minutes. The signal-to-noise ratio (SNR) in the continuum of each spectrum was typically 230 in echelle orders 61 through 70 of all spectra in the 2006 sequence, and 130 for the 2007 data which employed shorter exposures.

The spectra were extracted and numerical cross-correlation functions (CCF) were computed using the most recent version of the HARPS data-reduction software \citep{mayor2009harps}. This involves order-by-order cross-correlation of the extracted stellar spectrum with a weighted mask function, as described by \citet{baranne96elodie} and \citet{pepe2002harps}. The mask used for these observations  was derived from the line list for a star of spectral type K5. The SNR in the pseudo-continuum to either side of the cross-correlation peak was measured to be $\sim 6300$ for a typical spectrum in the 2006 series, and $\sim 3900$ in the 2007 sequence. The cross-correlation process thus yields a multiplex gain factor of approximately 29 in SNR, for both data sets, allowing the SNR in any individual CCF to be estimated directly from the SNR of the original spectrum.

\subsection{Stellar CCF profile model}

The CCF of a slowly-rotating star exhibits a flat continuum and a single dip, which can be fitted to high precision with a gaussian profile. We assume that the local line profile at any point on the surface of a limb-darkened, rotating star takes the form of a gaussian,
\begin{equation}
g(x)=\frac{1}{\sqrt{2\pi}s}\exp\left(-x^2/2s^2\right),
\end{equation}
where the dimensionless velocity $x=v/v\sin I$ and gaussian sigma $s=\sigma/v\sin I$ are expressed in units of the projected stellar equatorial rotation speed $v\sin I$. 

We assume a standard linear limb darkening model with limb-darkening coefficient $u$:
\begin{equation}
B(\mu)=B(1)(1-u+u\mu)
\label{eq:limblinear}
\end{equation}
using the standard convention that the direction cosine of the local surface normal relative to the line of sight is $\mu=\cos\theta$, with $\theta=0$ at the centre of the stellar disc. Assuming solid-body rotation, the limb-darkened rotation profile 
is obtained by substituting $\mu=\sqrt{1-x^2-y^2}$ and integrating Eq.~\ref{eq:limblinear} between limits $y=\pm\sqrt{1-x^2}$, where $x=v/v\sin I$ as before. The resulting expression is normalised by integrating over the range $-1<x<1$ to obtain \citep{gray1976oasp}
\begin{equation}
f(x)=\frac{6\,\left( \left( 1 - u \right) \,{\sqrt{1 - x^2}} - 
      \pi \,u\,\left( x^2 - 1 \right) /4 \right) }{\pi \,\left( 3-u \right) }.
\end{equation}
These expressions are normalised such that
\begin{equation}
\int_\infty^\infty g(x)dx=\int_{-1}^{1}f(x)dx=1.
\end{equation}
The observed rotation profile is the convolution of the two functions,
\begin{equation}
h(x) =\int_{-1}^1 f(z)g(x-z)dz,
\end{equation}
which is computed via a straightforward numerical integration. Since the CCFs are computed in the velocity frame of the solar-system barycentre, the shift of the model CCF to match a Keplerian orbit solution is achieved by defining $x_{ij}$ for a pixel $i$ with velocity $v_{ij}$ in the barycentric frame to be
\begin{equation}
x_{ij} = v_{ij}-(K(e\cos\omega+\cos(\nu_j+\omega))+\gamma)
\end{equation}
where $\nu_j$ is the true anomaly at the time of the $j$th observation and all other orbital elements $K$, $e$,$\omega$, $\gamma$ have their usual meanings.

The model of the $j$th observation of the stellar CCF is thus $h(x_{ij})$.

\begin{figure}
\includegraphics[width=7.5cm]{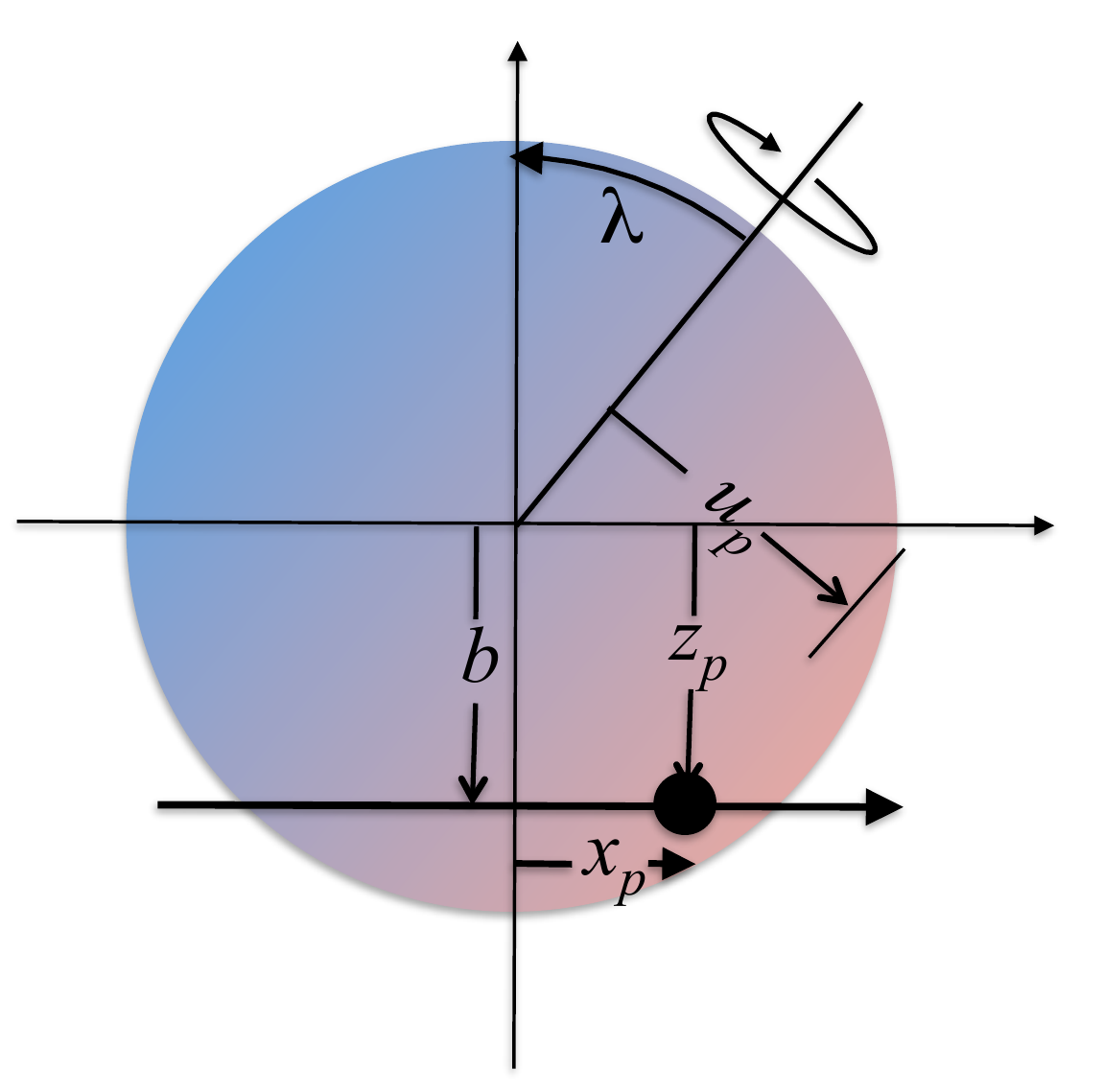}
\caption[]{Coordinate system used in definition of missing-light model. The instantaneous orbital coordinates of the planet, $x_p$ and $z_p$, are projected on the plane of the sky. The coordinate $u_p$ denotes projected distance from the stellar rotation axis. In this example, the projected obliquity $\lambda$ is a positive angle. At this obliquity, the planet's trajectory, with impact parameter $b$, carries it mainly across redshifted parts of the stellar surface, as denoted by the colour scale.}
\label{fig:cartoon}
\end{figure}

\subsection{Missing-light model during transit}

At any moment, the position of the planet on the plane of the sky is given by
\begin{eqnarray}
x_p&=&r\sin(\nu+\omega-\pi/2)\\
z_p&=&r\cos(\nu+\omega-\pi/2)\cos i
\end{eqnarray}
where $r$ is the instantaneous distance of the planet from the star, and $i$ is the inclination of the orbital angular momentum vector to the line of sight. The $z$ axis is parallel to the projected direction of the orbital axis (Fig.~\ref{fig:cartoon}). If $x_p$, $z_p$ and $r$ are expressed in units of the stellar radius, the fraction of the starlight blocked by the planet during the total part of a transit is 
\begin{equation}
\beta=\frac{R_p^2}{R_*^2}\frac{1-u+u\mu}{1-u/3}
\end{equation}
where $\mu = \sqrt{1-x_p^2-z_p^2}$.

\begin{figure}
\includegraphics[width=7.5cm,angle=270]{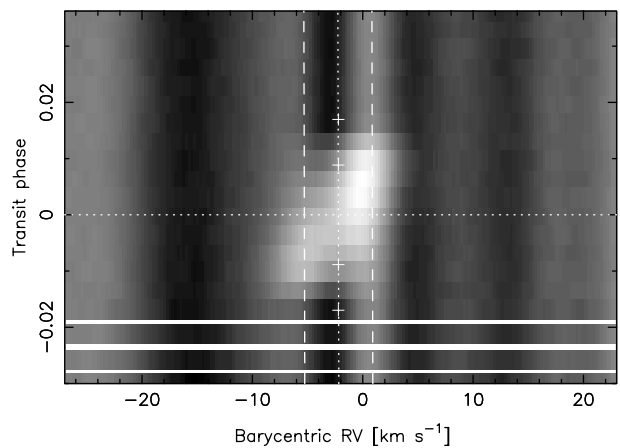}
\includegraphics[width=7.5cm,angle=270]{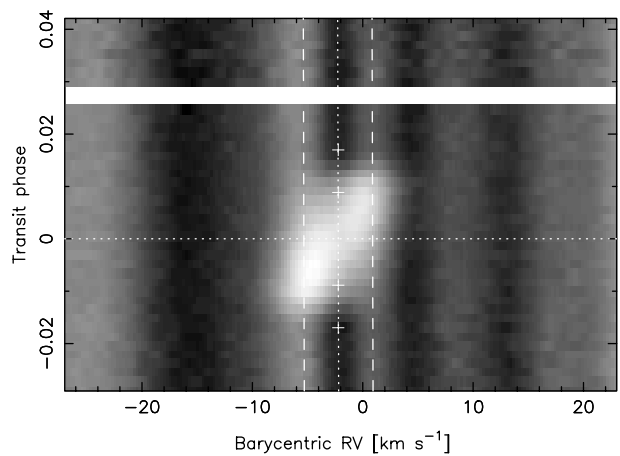}
\caption[]{Residual maps of the time-series CCFs for the night of 2006 September 7/8 (upper) and 2007 August 28/29 (lower). The model stellar profile $h_{ij}$ has been subtracted, leaving the bright travelling signature of the starlight blocked by the planet during the transit. This is seen superimposed on a fixed-pattern residual spectrum $\alpha_i$ which remains stable during each night but changes subtly from one night to the next.The horizontal dotted line gives the phase of mid transit. Crosses denote times of first, second, third and fourth contact. The stellar radial velocity is plotted as a dotted curve flanked by dashed lines at $\pm v\sin I$.}
\label{fig:bumpstripes}
\end{figure}

\begin{figure}
\includegraphics[width=7.5cm,angle=270]{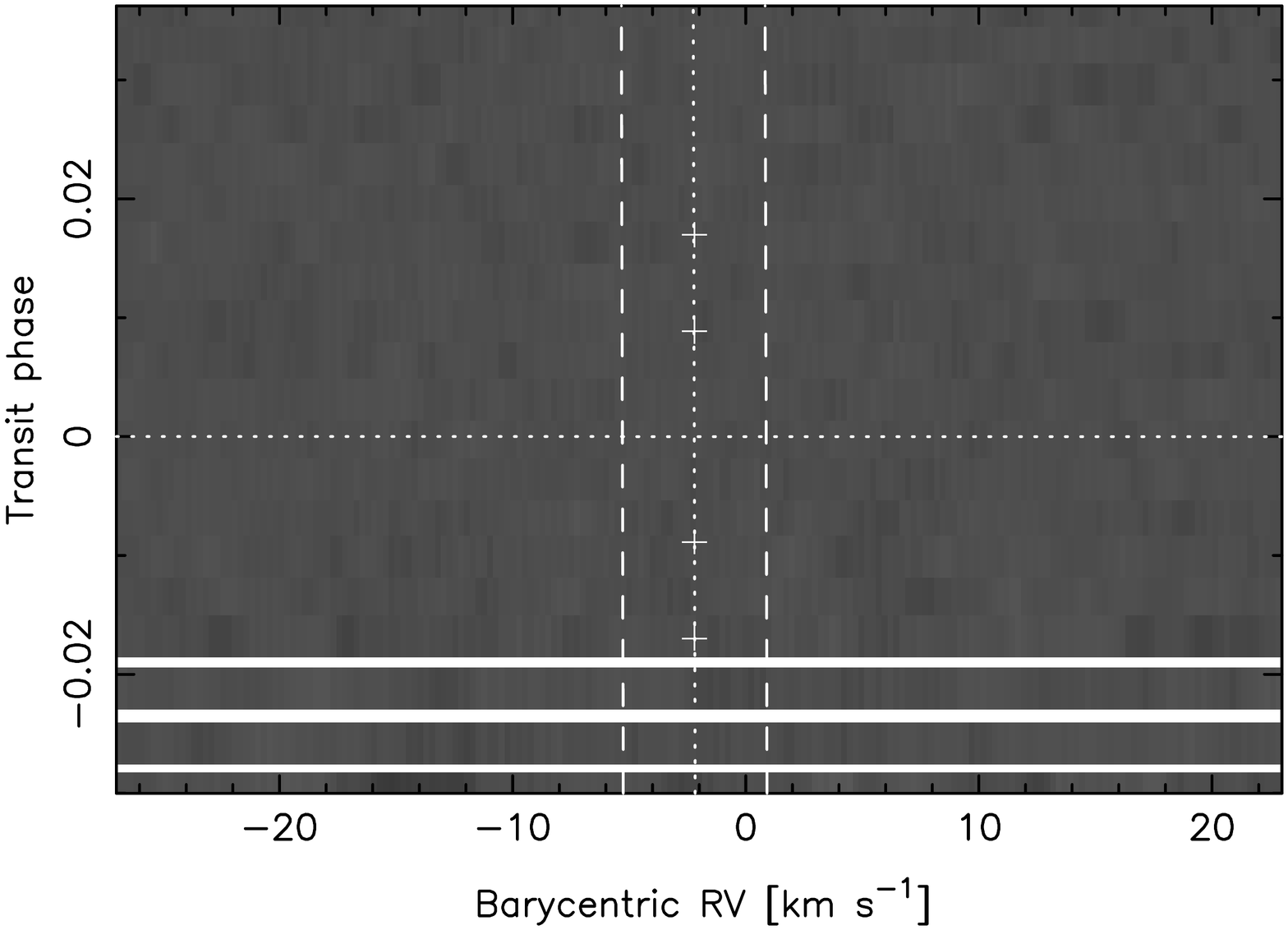}
\includegraphics[width=7.5cm,angle=270]{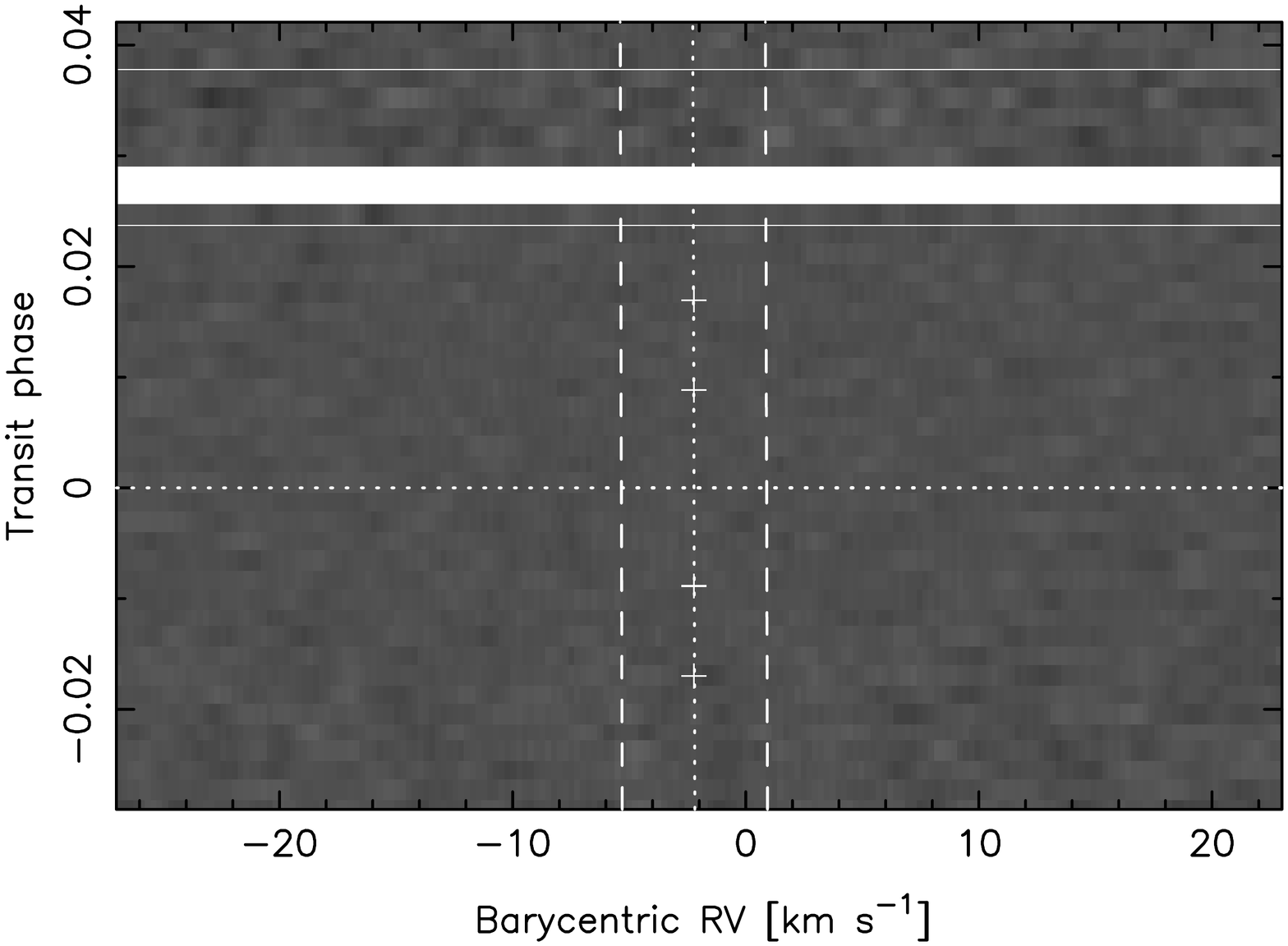}
\caption[]{Residual maps of the time-series CCFs for the night of 2006 September 7/8 (upper) and 2007 August 28/29 (lower). The scaled best-fit model $h_{ij}$ of the stellar profile, the travelling planet signature $\beta\ g(x_{ij}-u_p)$ and the  fixed sidelobe pattern $\alpha_i$ have all been subtracted from the data as in Eq.~\ref{eq:chisq}, leaving only noise.}
\label{fig:noise}
\end{figure}

The radial velocity of the missing starlight relative to the star's instantaneous velocity depends on the projected distance of the planet from the stellar rotation axis. Following \citet{winn2005rm} we define $\lambda=\phi_{\rm spin}-\phi_{\rm orbit}$ where $\phi$ denotes position angle in the plane of the sky.

Rotating the coordinate frame in the plane of the sky, we obtain the perpendicular distance of the planet from the stellar rotation axis in units of the stellar radius:
\begin{equation}
u_p=x_p\cos\lambda - z_p\sin\lambda,
\end{equation}
which in turn yields the barycentric velocity of the missing light, 
$v_p=u_p v\sin I$. 

The combined model of the stellar CCF and the missing starlight during the total part of the transit is thus
\begin{equation}
p_{ij}=h(x_{ij}) + \beta\ g(x_{ij}-u_p).
\end{equation}
Analytic expressions for computing $\beta$ and $u_p$ during the partial phases of the transit are given by \citet{ohta2005rm} and \citet{gimenez2006rm}.

\begin{table*}
\caption{System parameters from recent studies of HD 189733b. Values adopted from previous work are italicised.}
\label{tab:fixedparms}
\begin{tabular}{lccccr}
Parameter & \citet{winn2006hd189733} & \citet{pont2007hst} & \citet{triaud2009rm} & This study & Units\\
\hline\\
$T_c$ & $2453937.7759\pm 0.0001$ & $2453931.12048\pm 0.00002$ & $2453988.80339\pm 0.00006$ & {\it 2453988.8034} &d\\
$P$     & {\em 2.218575}     & $2.218581\pm 0.000002$  &  $2.218573\pm 0.000001$  & {\it 2.218573} &d\\
$K_s$ & -                   & -                        & $0.202\pm 0.001$     & {\it 0.202} &km s$^{-1}$\\
$e$     & 0                  & 0                       & $0.0041\pm 0.0022$ & {\it 0.0041} & \\
$\omega$ & -             & -                        & $-24\pm 34$              & {\it -24} &degrees\\
$b$     & -                  & $0.671\pm 0.008$ & $0.687\pm .006$       & {\it 0.687} &\\
$i$       & $86.1\pm 0.2$ & $85.68\pm 0.04$ & $85.51\pm 0.1$         & {\it 85.51} &degrees\\
$M_*$ & ${\it 0.82\pm 0.03}$ & ${\it 0.82\pm 0.03}$  & ${\it 0.82\pm 0.03}$         & {\it 0.82} &M$_\odot$\\
$R_*$  & $0.73\pm 0.02$ &  $0.75\pm0.01$ & $0.77\pm 0.01$         & {\it 0.77} &R$_\odot$\\
$R_p/R_*$ & - & $0.1572\pm 0.0004$      & $0.1581\pm 0.0005$ & {\it 0.1581} &\\
\end{tabular}
\end{table*}

The term $\beta\ g(x_{ij}-u_p)$ represents the travelling gaussian ``bump" produced by the blocked starlight. The appearance of this residual bump after subtraction of the model stellar profile from the observed CCFs is illustrated in greyscale form in Fig.~\ref{fig:bumpstripes}. The coefficient $\beta$ depends on the ratio $R_p/R_*$ and the limb-darkening coefficient $u$. The width of the gaussian $g$ gives an additional constraint on the non-rotating local CCF width $s$, given the reasonable assumption that $R_p<< R_*$. The trajectory of the planet signature through the line profile is described by $u_p$, which in turn depends on $v \sin I$ and $\lambda$. These separate constraints on $s$, $u$ and $v\sin I$ are valuable, since they help to break any degeneracy in the description of the shape of the stellar CCF, particularly when $v\sin I$ is small compared to the intrinsic linewidth.

\subsection{Model fitting and error scaling}

The data and model are both orthogonalised by subtracting their optimal mean values:
\begin{eqnarray}
d'_{ij}&=&d_{ij}-(\sum_{ij} d_{ij} w_{ij}/\sum_{ij} w_{ij})\\
p'_{ij}&=&p_{ij}-(\sum_{ij} p_{ij} w_{ij}/\sum_{ij} w_{ij}).
\end{eqnarray}
using inverse-variance weights $w_{ij}=1/\sigma^2_{ij}$.
The multiplicative constant $\hat{A}$ by which the model must be scaled 
to give an optimal fit to the data is then given by the expression
\begin{equation}
\hat{A}=\sum_{ij} d'_{ij} p'_{ij} w_{ij} / \sum_{ij} {p'}_{ij}^2 w_{ij}.
\end{equation}

The CCF obtained by cross-correlating a late-type stellar spectrum with a line mask inevitably contains weak sidelobes, caused by random alignments between stellar lines and mask lines at arbitrary shifts. The cross-correlation dip therefore appears superimposed on a background pseudo-continuum which exhibits weak fixed-pattern residuals. These appear as an alternating pattern of light and dark vertical stripes in Fig.~\ref{fig:bumpstripes}. Imperfect modelling and removal of the spectrograph blaze function prior to  cross-correlation can also produce a weak asymmetry in the CCF profile, which would leave a similar pattern in the residuals. At this stage in the calculation, the sidelobe pattern is removed
by computing and subtracting the optimal average of the residual spectra:
\begin{equation}
\alpha_i = \sum_j(d'_{ij}-\hat{A}p'_{ij})w_{ij}/\sum_j w_{ij}.
\end{equation}
The sidelobe pattern can also be influenced by moonlight and telluric absorption levels, and potentially by weak, time-variable asymmetries in the CCF caused by starspots. This produces subtle changes in the sidelobe pattern from one night to the next. Little change is seen within a single night. The method is therefore most effective when the spectrum-averaging summations are performed separately for different nights. The residuals following the subtraction of the sidelobes and the signature of the light blocked by the planet are shown in Fig.~\ref{fig:noise}.

The goodness of the model fit to the observed CCF is then measured via the statistic
\begin{equation}
\chi^2=\sum_{i=1}^n (d'_{ij}-\hat{A}p'_{ij}-\alpha_i)^2w_{ij}.
\label{eq:chisq}
\end{equation}
Care is needed when estimating the error bars on both the data and the fitted parameters. The spectral resolving power of HARPS is $R=\lambda/\Delta\lambda=120000$, yielding a velocity resolution of 2.5 km~s$^{-1}$. The CCFs produced by the HARPS data reduction pipeline are oversampled, being binned in velocity increments of 0.25 km~s$^{-1}$. The errors on adjacent points in the CCF are therefore strongly correlated. In order to ensure that all data points contributing to $\chi^2$ are statistically independent, we bin the data and the fitted model in velocity, by a factor ten before computing $\chi^2$. Initially we estimate the relative uncertainties $\sigma_{ij}$  of the CCF data in each spectrum from the SNR of the original spectrum using the multiplex gain factor described in Section~\ref{sec:obsred} above.

\subsection{Markov-chain Monte Carlo parameter fitting}

Given a sequence of CCFs derived from a set of spectra densely sampled during transit and distributed around the rest of the orbit, the overall $\chi^2$ statistic is therefore a function of the orbital parameters $K$, $e$, $\omega$ and $\gamma$; the orbital period $P$ and the epoch $T_p$ of periastron; the parameters $v\sin I$, $s$ and $u$ describing the shape of the rotationally-broadened stellar CCF and the continuum tilt; and the error scaling factor $f$.

The limited phase range of the HARPS observations does not permit as reliable a determination of the full set of orbital parameters as the combined SOPHIE and HARPS data used by \citet{triaud2009rm}. We therefore chose not to fit the full radial-velocity orbit and the photometric transit parameters simultaneously with the spectroscopic transit parameters. Instead, we fixed the values of all other model parameters at the values given by \citet{triaud2009rm} in their recent combined photometric and spectroscopic analysis of the same system. These values are summarised and compared with their values from earlier studies in Table~\ref{tab:fixedparms}. Only the system centre-of-mass velocity, $\gamma$, was re-determined directly from the data. This was done because the star is known to exhibit starspot activity, so small night-to-night changes in $\gamma$ are to be expected as the star rotates. Although spot activity could in principle affect the apparent value of $K$ too, the latter is well-determined from RV observations collected over many orbits. Any degradation of the fit by starpot-induced changes in the apparent radial acceleration during the transit is expected to be small.

We determined the model parameters $v\sin I$,  $\lambda$, $s$ and $\gamma$ and their joint posterior probability distribution using a Markov-chain Monte Carlo (MCMC) approach.

\begin{figure*}
\includegraphics[width=10cm,angle=270]{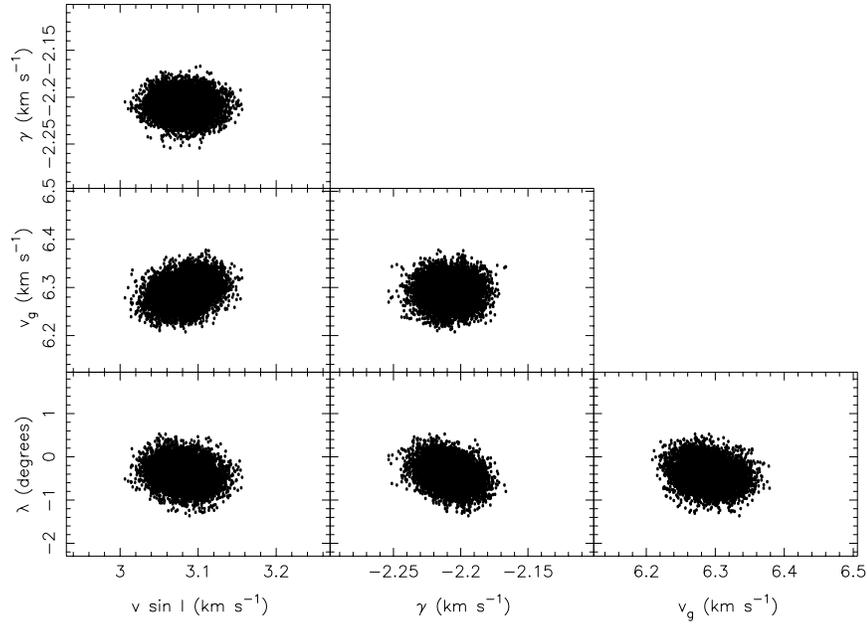}
\caption[]{Correlation diagrams for the joint posterior probability distributions of the four model fitting parameters $\{v\sin I, \lambda, v_{\rm g}, \gamma\}$ for the transit of HD 189733b observed on 2006 September 7/8.}
\label{fig:plotmatrix_2006}
\end{figure*}

\begin{figure*}
\includegraphics[width=10cm,angle=270]{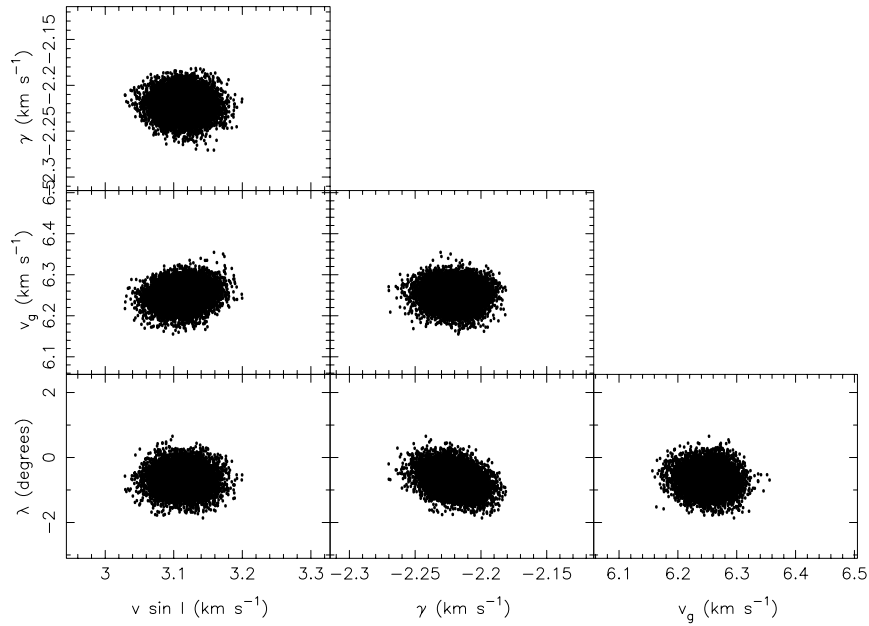}
\caption[]{Correlation diagrams for the joint posterior probability distributions of the four model fitting parameters $\{v\sin I, \lambda, v_{\rm g}, \gamma\}$ for the transits of HD 189733b observed on 2007 August 28/29.}
\label{fig:plotmatrix_2007}
\end{figure*}

Strongly-correlated pairs of parameters are best avoided in MCMC, since they lead to slow convergence and undesirably strong correlations between successive steps in the chain \citep{tegmark2004mcmc}.  We expect the parameters $s$ and $v\sin I$, which together determine the  width of the stellar CCF outside transit, to be correlated to some extent, since their quadrature sum is fixed by the width of the observed CCF. To maintain orthogonality between jump parameters we replaced $s$ with the full width at half maximum (FWHM) of the CCF, which is approximated by
\begin{equation}
v_{\rm ccf} = [ (v\sin I)^2 + v_{\rm g}^2]^{1/2}=v\sin I [ 1 + 4s^2\ln 2]^{1/2}.
\end{equation}
Here $v_{\rm g}$ is the FWHM of the gaussian representing the local stellar and instrumental profile:
\begin{equation}
v_{\rm g} = 2s\ v\sin I\sqrt{\ln 2}.
\end{equation}

The Markov chain is constructed by repeatedly computing $\chi^2$ for a sequence of parameter sets $\{v\sin I, \lambda, v_{\rm ccf}, \gamma\}$. 
At each step $k$, a set of trial parameters is computed by applying gaussian perturbations to the preceding set:
\begin{eqnarray*}
 v\sin I_k &=& v\sin I_{k-1} + f\sigma_{v\sin I}G(0,1)\nonumber\\
\lambda_k &=& \lambda_{k-1} + f\sigma_{\lambda}G(0,1)\nonumber\\
v_{{\rm ccf},k} &=& v_{{\rm ccf},k-1} + f\sigma_{v_{\rm ccf}}G(0,1)\nonumber\\
\gamma_k &=& \gamma_{k-1} + f\sigma_{\gamma}G(0,1)\nonumber
\end{eqnarray*}
The scale factor $f$ is of order unity. The goodness-of-fit
is recalculated for the new parameter set, and the change $\Delta\chi^2=\chi^2_k-\chi^2_{k-1}$ is computed. 
The new set is either accepted or rejected according to the Metropolis-Hastings rule: if $\Delta\chi^2 < 0$ the set is accepted,
and if $\Delta\chi^2 > 0$, the new step is accepted with probability $e^{-\Delta\chi^2/2}$. If the proposed parameters are accepted, they are written to the next record in the Markov chain. If not, the previous set of values is recorded. The scale factor $f$ is tuned to ensure
that roughly 25 percent of all steps are accepted once the chain has settled down to a steady state; by trial-and-error we found that
$f=0.7$ achieved the desired acceptance rate. 

The chain evolved to a steady state within about 100 accepted proposals. We terminated the initial ``burn-in" phase after at least 100 proposals had been accepted, when the value of $\chi^2_k$ exceeded the median $\chi^2$ of all previous steps for the first time \citep{knutson2008spitzercal}. At this stage we refined our estimates of the variances of the binned CCF data, replacing the initial estimates $\sigma^2_{ij}$ with $\sigma^2_{ij}\chi^2_{{\rm min},j}/n_d$. Here $\chi^2_{{\rm min},j}$ is the contribution of the $j$th spectrum to the $\chi^2$ statistic for the best-fitting model in the chain so far, and $n_d$ is the number of degrees of freedom associated with the spectrum. This ensures that $\chi^2$ is scaled correctly, and that the posterior probability distributions of the fitted parameters have the correct variances.

We then ran the chain for a further 100 proposal acceptances, and re-evaluated the variances of the four fitting parameters from the chains themselves. After a further short burn-in period, we performed a production run of $10^4$ accepted proposals. The correlation lengths of all four chains were computed as described by \citet{tegmark2004mcmc}, and found to be of order 10 steps in all four parameters.

The profile decomposition method carries higher computational overheads than the anomalous RV approaches developed by \citet{winn2005rm} and \citet{hirano2009rm}. Substantial speed gains can be realised by performing the numerical integration to generate the synthetic stellar profile only once per step in the Markov chain, and shifting the profile to the appropriate orbital velocity at the time of each observation. The MCMC analyses presented here, with chains comprising $10^4$ accepted steps, incurred execution times of order 50 minutes on a standard 2.66 GHz laptop processor, as opposed to a few minutes for a conventional analysis.

\section{Results}

In Figs.~\ref{fig:plotmatrix_2006} and \ref{fig:plotmatrix_2007} we show the correlation diagrams for all four jump parameters, for the nights  of 2006 September 7/8 and 2007 August 28/29. In these plots we show the FWHM $v_{\rm g}$ of the intrinsic gaussian profile, rather than the FWHM $v_{\rm ccf}$ of the convolved profile. Contrary to our initial expectations, we find that $v\sin I$ and $v_{\rm g}$ are only very weakly correlated. This confirms that the two independent measures of both quantities (from the shape of the stellar profile and the width and trajectory of the missing-light signature) serve to fix their values unambiguously.

The four fitted parameters for each of the two transits are listed in Table~\ref{tab:fitparms}, for three different values of the linear limb-darkening coefficient $u$. These values bracket the value $u\simeq 0.725$ from the {\sc atlas9} models of \citet{claret2000} for a stellar effective temperature of 5050K, $\log g= 4.5$ and near-solar metallicity, as reported by \citet{bouchy2005} for HD 189733. The tabulated values illustrate that the best-fitting values of $v\sin I$ and $v_{\rm g}$ are only very weakly dependent on the precise value of $u$.

\begin{table}
\caption[]{Mean and one-sigma errors for MCMC model parameters fitted to the transit events recorded on 2006 September 7/8 and 2007 August 28/29. Results are shown for three MCMC analyses using different values of the linear limb-darkening coefficient $u$.}
\label{tab:fitparms}
\begin{tabular}{lccr}
\hline\\
Parameter & 2006 Sept 7/8  & 2007 Aug 28/29& Units\\
\hline\\
$u=0.675$&&\\
$v \sin I$ &   $3.071 \pm   0.019$ & $3.113 \pm  0.0252$&km s$^{-1}$\\
$\lambda$  &  $-0.37 \pm   0.234$ & $ -0.65\pm 0.32$&degrees\\
$v_{\rm g}$ &   $6.277 \pm  0.024$ & $6.243 \pm  0.026$&km s$^{-1}$\\
$\gamma$ &  $-2.209 \pm  0.010$ & $ -2.223 \pm  0.012$&km s$^{-1}$\\
& & \\
$u=0.725$&&\\
$v \sin I$ &   $3.0835 \pm   0.020$ & $3.115 \pm  0.024$&km s$^{-1}$\\
$\lambda$  &  $-0.40 \pm   0.24$ & $ -0.65\pm 0.33$&degrees\\
$v_{\rm g}$ &   $6.286 \pm  0.022$ & $6.248 \pm  0.026$&km s$^{-1}$\\
$\gamma$ &  $-2.209 \pm  0.011$ & $ -2.221 \pm  0.013$&km s$^{-1}$\\
& & \\
$u=0.775$&&\\
$v \sin I$ &   $3.088 \pm   0.021$ & $3.126 \pm  0.024$&km s$^{-1}$\\
$\lambda$  &  $-0.35 \pm   0.25$ & $ -0.67\pm 0.33$&degrees\\
$v_{\rm g}$ &   $6.293 \pm  0.024$ & $6.252 \pm  0.025$&km s$^{-1}$\\
$\gamma$ &  $-2.208 \pm  0.012$ & $ -2.222 \pm  0.013$&km s$^{-1}$\\
\hline\\
\end{tabular}
\end{table}

Although the two sets of observations were made nearly a year apart, the fitted values of the four parameters agree within their uncertainty ranges. The radial velocity $\gamma$ of the system's centre of mass remained constant to within 10~m~s$^{-1}$. 
The spin-orbit misalignment angle $\lambda$ is found to be slightly negative in both years. We find an absolute value slightly, but not significantly, smaller than the $(0.85\pm 0.30)^\circ$ measured by Triaud et al. The uncertainty in $\lambda $ is also 0.3 degrees, comparable with the uncertainty derived by \citet{triaud2009rm} from the same data. 

Our value of  $v\sin I=3.10\pm 0.03$ km s$^{-1}$ is significantly lower than the $3.32^{+0.02}_{-0.07}$ km s$^{-1}$ obtained by \citet{triaud2009rm} using an uncorrected fit to the HARPS radial velocities. As  Triaud et al. noted, the radial-velocity measurements obtained during transit by gaussian fitting show systematic departures from the intensity-weighted average velocity of the unobscured parts of the stellar disc. They modelled the departure of the radial-velocity centroid from the gaussian-fitting velocity as a function of $v\sin I$, and found that  $v\sin I = 3.05$ km s$^{-1}$ yielded the best fit to the observed pattern of residuals. An identical pattern of residuals was seen in the earlier R-M study of this system by \citet{winn2006hd189733}. Their empirical correction procedure yielded a lower $v\sin I = 2.97\pm 0.22$ km s$^{-1}$. Our direct tracking of the missing-starlight component in the CCF eliminates the need for such empirical corrections to indirect measures of the missing-light velocity, and yields a more precise result that lies within the error ranges of both studies.

\section{Discussion and conclusions}

In principle we could determine a lower limit on the stellar radius from $v\sin I$ and the photometric rotation period, which was determined by   \citet{henry2008hd189733} to be $11.953\pm 0.009$ days. The very close alignment of the projected orbital and spin axes in the plane of the sky makes it highly probable that the inclination $I$ of the stellar spin axis to the line of sight should be very close to the inclination $i=85.5$ degrees of the planet's orbit. Assuming $\sin I = \sin i = 0.997$, our value for $v\sin I$ yields a direct estimate of the stellar radius $R_*=0.732\pm 0.007$ R$_\odot$, independently of the usual assumptions involving isochrone fits to the stellar density and effective temperature. This is significantly smaller than the $R_*=0.77\pm 0.01$ R$_\odot$ determined by \citet{triaud2009rm}, and marginally less than the $R_*=0.75\pm 0.01$ R$_\odot$ found by \citet{pont2007hst}. 

We suspect that the discrepancy arises from differential rotation of the stellar photosphere. \citet{gaudi2007rm} 
pointed out that under some circumstances, the RM effect could be used to measure stellar differential rotation. As with most previous studies of this kind, we have measured $v\sin I$ under the assumption that the star rotates as a solid body. In all stars for which differential rotation has been measured directly, however, a solar-like pattern is found: the equator rotates faster than the poles, with the departure from the equatorial spin rate having a sine-squared dependence on latitude. The typical difference in rotational angular frequency $\Delta\Omega = \Omega_{\rm equator}-\Omega_{\rm pole}$ is found to be of order 0.04 radian per day for rapidly-rotating stars K dwarfs similar to HD 189733, and only weakly dependent on rotation rate \citep{barnes2005diffrot}. This is very similar to the solar value of $\Delta\Omega$.

If the starspots giving rise to the photometric modulation signal are concentrated at the same stellar latitude as the planet's path across the stellar disc, we should derive the correct value of the stellar radius. Unfortunately we have no way of determining the dominant latitude from which the modulation signal originates. If the orbital and spin axes are closely aligned, the impact parameter of the planet's path across the star carries it across a stellar latitude of 39$^\circ$ in the hemisphere facing away from the observer. Although \citet{pont2007hst}  noted the passage of the planet across one or two isolated dark spots, these spots are significantly foreshortened. It is reasonable to expect the modulation signal to arise mainly from spots at intermediate to low latitude in the observer's hemisphere. A difference in latitude of 20 or 30 degrees is sufficient to give a difference in angular velocity $\delta\Omega\simeq 0.015$ radians per day between the main spot belts and the latitudes traversed by the planet. This discrepancy yields a stellar radius that is too small by about 3 percent, which is sufficient to reconcile our radius estimate with those of \citet{pont2007hst} and \citet{triaud2009rm}.

We conclude that direct modelling of the cross-correlation function during a planetary transit is feasible even for a host star such as HD~189733 whose $v\sin I$ is comparable to the intrinsic line width. The profile decomposition method described here yields values and error estimates for the stellar spin rate and orbital obliquity that agree closely with the results of previous studies of this system. The parameter values are free of the systematic errors that occur when velocity measurements are derived from gaussian fits to line profiles with inherent time-variable asymmetry. This circumvents the need for semi-empirical corrections when modelling the Rossiter-McLaughlin effect. The method makes the fullest use of all the information present in the cross-correlation function: $v\sin I$ is tightly constrained by the shape of the stellar rotation profile as well as by the spectral signature of the starlight blocked by the planet. The stellar radius obtained from the photometric rotation period and the stellar $v\sin I$ agrees well with values obtained from high-precision transit photometry, if a modest amount of differential rotation is present.

The method described here could in principle be extended to observations for which an iodine cell is used to track the spectrograph PSF and wavelength scale. The analysis of such observations requires a reference spectrum of the host star, taken with the same instrument without the iodine cell. An artificial reference spectrum could be generated by convolving a limb-darkened rotation profile with the the spectrum of a narrow-lined star of the same spectral type. The Doppler signature of the light blocked by the planet could be mimicked by scaling, shifting and subtracting the same narrow-lined spectrum from the broadened stellar spectrum; the result would then be used as the reference spectrum. A very similar procedure was described by \citet{winn2005rm} for calibrating the departure of the velocities measured during a transit of HD 209458b from the predictions of the \citet{ohta2005rm} model, using high-resolution solar spectra. 

\section*{Acknowledgments}

We thank the HARPS consortium for providing some of the observations presented here, and the ESO La Silla staff for their support at the telescope. 
This research has made use of the ESO Data Archive, the NASA Astrophysics Data System, and the VizieR catalogue access tool, CDS, Strasbourg, France.
AHMJT and GM  acknowledge studentship support from the Swiss Fond National de Recherche Scientifique and the UK Science and Technology 
Facilities Council respectively. We thank the referee, Dr J. Winn, for his insightful suggestions for improving the original manuscript.

\bibliographystyle{mn2e}

\bsp

\label{lastpage}

\end{document}